\documentclass[reqno,12pt]{amsart}
\usepackage{amsmath}
\usepackage{amsxtra}
\usepackage{amscd}
\usepackage{amsthm}
\usepackage{amsfonts}
\usepackage{amssymb}
\usepackage{eucal}
\usepackage[matrix,arrow,curve]{xy}
\usepackage{color}
\usepackage{xcolor}
\usepackage{cite}
\theoremstyle{plain}
\newtheorem{Thm}[subsection]{Theorem}
\newtheorem{Cor}[subsection]{Corollary}
\newtheorem{Lem}[subsection]{Lemma}
\newtheorem{Prop}[subsection]{Proposition}
\newtheorem{Conj}[subsection]{Conjecture}

\theoremstyle{definition}
\newtheorem{Def}[subsection]{Definition}

\theoremstyle{remark}

\newtheorem{Rem}[subsection]{Remark}

\newtheorem{conj}{Conjecture}

\numberwithin{equation}{section}
\renewcommand{\rm}{\normalshape}

\newenvironment{thm}[1]%
  { \begin{Thm} \label{T:#1} \ifShowLabels \TeXref{T:#1} \fi }%
  { \end{Thm} }

\renewcommand{\th}[1]{\begin{thm}{#1} \sl }
\renewcommand{\eth}{\end{thm} }

\newenvironment{lemma}[1]%
  { \begin{Lem} \label{L:#1} \ifShowLabels \TeXref{L:#1} \fi }%
  { \end{Lem} }
\newcommand{\lem}[1]{\begin{lemma}{#1} \sl}
\newcommand{\elem}{\end{lemma}}

\newenvironment{propos}[1]%
  { \begin{Prop} \label{P:#1} \ifShowLabels \TeXref{P:#1} \fi }%
  { \end{Prop} }
\newcommand{\prop}[1]{\begin{propos}{#1}\sl }
\newcommand{\eprop}{\end{propos}}

\newenvironment{corol}[1]%
  { \begin{Cor} \label{C:#1} \ifShowLabels \TeXref{C:#1} \fi }%
  { \end{Cor} }
\newcommand{\cor}[1]{\begin{corol}{#1} \sl }
\newcommand{\ecor}{\end{corol}}

\newenvironment{defeni}[1]%
  { \begin{Def} \label{D:#1} \ifShowLabels \TeXref{D:#1} \fi }%
  { \end{Def} }
\newcommand{\defe}[1]{\begin{defeni}{#1} \sl }
\newcommand{\edefe}{\end{defeni}}

\newenvironment{remark}[1]%
  { \begin{Rem} \label{R:#1} \ifShowLabels \TeXref{R:#1} \fi }%
  { \end{Rem} }
\newcommand{\rem}[1]{\begin{remark}{#1}}
\newcommand{\erem}{\end{remark}}

\newenvironment{conjec}[1]%
  { \begin{Conj} \label{Co:#1} \ifShowLabels \TeXref{Co:#1} \fi }%
  { \end{Conj} }
\renewcommand{\conj}[1]{\begin{conjec}{#1} \sl }
\newcommand{\econj}{\end{conjec}}

\newcommand{\eq}[1]%
  { \ifShowLabels \TeXref{E:#1} \fi
    \begin{equation} \label{E:#1} }
\newcommand{\eeq}{ \end{equation} }

\newcommand{\prf}{ \begin{proof} }
\newcommand{\epr}{ \end{proof} }

\def\be{\begin{equation}}
\def\ee{\end{equation}}
\def\ba{\begin{array}}
\def\ea{\end{array}}

\newif\ifShowLabels
\ShowLabelstrue
\newdimen\theight
\def\TeXref#1{%
  \leavevmode\vadjust{\setbox0=\hbox{{\tt
    \quad\quad {\small \rm #1}}}%
  \theight=\ht0
  \advance\theight by \lineskip
  \kern -\theight \vbox to
  \theight{\rightline{\rlap{\box0}}%
  \vss}%
  }}%
\ShowLabelsfalse% comment this out if labels should be printed

\numberwithin{equation}{section} \makeatletter
\@addtoreset{equation}{section}

\makeatletter
\g@addto@macro{\endabstract}{\@setabstract}
\newcommand{\authorfootnotes}{\renewcommand\thefootnote{\@fnsymbol\c@footnote}}%
\makeatother

\begin{document}

\begin{flushright}
FIAN-TD-2015-03 \\
\end{flushright}

\bigskip
\bigskip
\begin{center}
{ \large\textbf{Matrix model approach to minimal Liouville gravity \\ \vspace{3mm}revisited
 }}
 \par
 \vspace{.8cm}

 \normalsize
 \authorfootnotes
V.~Belavin%\footnote{belavin@lpi.ru}
\textsuperscript{1,2,3}
and Yu.~Rud %\footnote{juju\_mbox@rambler.ru}
\textsuperscript{4,5}\par \bigskip

\begin{tabular}{ll}
$\qquad ^{1}$~\parbox[t]{0.9\textwidth}{\normalsize\raggedright
\small{I.~E. Tamm Department of Theoretical Physics, \,\,P.~N. Lebedev Physical \\Institute, Leninsky prospect 53, 119991 Moscow, Russia}}\\
$\qquad^{2}$~\parbox[t]{0.9\textwidth}{\normalsize\raggedright
\small{Department of Quantum Physics, Institute for Information Transmission Problems, Bolshoy Karetny per. 19, 127994 Moscow, Russia}}\\
$\qquad^{3}$~\parbox[t]{0.9\textwidth}{\normalsize\raggedright
\small{Department of Theoretical Physics, National Research Nuclear University MEPhI,
Kashirskoe shosse 31, 115409 Moscow, Russia}}\\
\end{tabular}\\
$\qquad^{4}$~\parbox[t]{0.9\textwidth}{\normalsize\raggedright
\small{L.~D. Landau Institute for Theoretical Physics, 142432 Chernogolovka, Russia}}\\
$\qquad^{5}$~\parbox[t]{0.9\textwidth}{\normalsize\raggedright
\small{Moscow Institute of Physics and Technology, 141700 Dolgoprudny, Russia}}
% \today
\end{center}
\vspace{.8cm}

\begin{abstract}
Using the connection with the Frobenius manifold structure, we study the
matrix model description of minimal Liouville gravity (MLG) based on the
Douglas string equation. Our goal is to find an exact discrete formulation
of the $(q,p)$ MLG model that intrinsically contains information about the
conformal selection rules. We discuss how to modify the Frobenius manifold
structure appropriately for this purposes. We propose a modification of the 
construction for Lee--Yang
series  involving the $A_{p-1}$ algebra instead of the
previously used $A_1$ algebra. With the new prescription, we calculate
correlators on the sphere up to four points and find full agreement with the
continuous approach without using resonance transformations.
\end{abstract}
\bigskip

\section{Introduction}
\label{secIntroduction}

Minimal Liouville gravity (MLG) is a special model of Liouville gravity~\cite{Polyakov:1981rd}
with the matter sector represented by the $(q,p)$ minimal CFT
model.\footnote{In this paper, we focus on the $A$-series of Virasoro minimal
models.} MLG is a BRST theory with the well-known structure of the Hilbert
space~\cite{Belavin:2006kv}. One of the main problems in the theory is to
compute the correlators of the primary cohomologies. Because of the
integrals over moduli involved in the construction of the correlators, this
computation requires quite sophisticated techniques~\cite{Zam:2004}. Only
expressions up to four-point correlation numbers have been found explicitly so
far~\cite{Zam:2005,Belavin:2006kv}. Here, we study an alternative approach
to MLG that gives a simple procedure for computing the correlation functions.
This approach is connected with the matrix model (MM)~\cite{Kazakov:1985ea,Kazakov:1986hu,Kazakov:1989bc,Staudacher:1989fy,Brezin:1990rb,Douglas:1989ve,Gross:1989vs,Krichever:1992sw} of two-dimensional
quantum gravity and also called a discrete approach. The basic fact about
the MLG--MM correspondence is the coincidence of the spectra of the
gravitational dimensions~\cite{Knizhnik:1988ak}. This result represents the
main support of the idea that the two approaches describe the same quantum
theory. In the $(q,p)$ model\footnote{We recall that $q$ and $p$ are two
coprime integers.} of MLG, the primary BRST cohomologies $O_{mn}$ are
labeled by two integers $m=1,\dots,q-1$ and $n=1,\dots,p-1$. In what
follows, we assume $q<p$. Hence, the main object of our study is the
generating function of the correlators of the primary cohomologies
\be
Z=\langle\exp\sum_{n}\lambda_{mn}O_{mn}\rangle_{\text{MLG}}=
\sum_{N=0}^{\infty}\sum_{n_i}\frac{\lambda_{m_1n_1}\ldots
\lambda_{m_Nn_{N}}}{N!}\langle O_{m_1n_1}\ldots O_{m_Nn_N}\rangle_{\text{MLG}}\,.
\label{Zcont}
\ee
The brackets $\langle\dots\rangle_{\text{MLG}}$ denote the integrated
correlation functions, which we therefore call correlation numbers. In what
follows, we call the parameters $\lambda_{mn}$ Liouville coupling constants.

As was first shown in the seminal KPZ paper~\cite{Knizhnik:1988ak}, the
scaling properties of the correlators in MLG are governed by the following
rule. The dependence of the correlator $G=\langle O_{m_1n_1}\ldots
O_{m_Nn_N}\rangle$ on the cosmological constant is given by
\be
G(\mu)=\mu^{\frac{p+q}{q}-\sum_{i=1}^N\delta_{m_in_i}}G(1)\,,
\ee
where
\be
\delta_{mn}=\frac{p+q-|pm-qn|}{2q},\quad O_{mn}\sim\mu^{-\delta_{mn}}\,.
\label{delta_mn}
\ee
Therefore, the first basic requirement for the dual approach is to reproduce
this spectrum. In the dual description of the $(q,p)$ MLG model, there are
two basic elements. For the spherical topology, we introduce the polynomial
\be
Q(y)=y^q+u_1y^{q-2}+\ldots+u_{q-1}\,,
\ee
where $y$ is an auxiliary variable (we discuss in more detail below, that
this polynomial defines the structure of a special Frobenius manifold (FM)
and the set $\{u_{\alpha}\}$ represents a special choice of coordinates on
this manifold)~\cite{Dijkgraaf:1990dj,Dubrovin:1992dz} and the so-called action
\be
S(t_{mn})=\underset{y=\infty}{\text{res}}\left(Q^{\frac{p+q}{q}}+
\sum_{m,n}^{pm-qn>0}t_{mn} Q^{\frac{pm-qn}{q}}\right),
\ee
which defines the generating function of the correlation numbers and appears
to be the subject of the string equation~\cite{Douglas:1989dd}. The
parameters $t_{mn}$ are known as KdV times (or couplings). Weights
$\tilde{\delta}_{mn}$ ($\tilde{\delta}_{11}=1$) can be assigned to the
couplings $t_{mn}$ and $y$ so that $Q(y)$ becomes quasihomogeneous of
weight 1/2. With the identification $t_{11}\sim\mu$, it can be easily
verified~\cite{Douglas:1989dd} that the spectrum of gravitational dimensions
is exactly the spectrum~\eqref{delta_mn} that appears in the continuous
approach to MLG, i.e., $\delta_{mn}=\tilde{\delta}_{mn}$. We thus obtain a
natural identification between the couplings of the two approaches
$t_{mn}=\lambda_{mn}$.

After Douglas had shown~\cite{Douglas:1989dd} that the scaling dimensions of
KdV times coincide with those of MLG coupling constants, there were attempts
to verify the correspondence between the two approaches at the level of the
correlation functions. However, the obtained correlators failed to
satisfy conformal selection rules already at one-point level. A possible
resolution of this problem was formulated by Moore, Seiberg, and
Staudacher~\cite{Moore:1991ir}. The idea was based on the observation that
MLG by definition contains an ambiguity related to the fact that the
correlation numbers, which are given by integrals over moduli spaces, depend
on the contact behavior when the positions of a few insertions collide with
each other. Indeed, such ultraviolet information is not provided in the
standard CFT formalism based on the notion of the operator product
expansion. This ambiguity allows extending the possible form of the
relations between Liouville coupling constants and KdV times,
\be
t_{mn}=\lambda_{mn}+\sum_{m_1n_1m_2n_2}
A_{mn}^{m_1n_1,m_2n_2}\lambda_{m_1n_1}\lambda_{m_2n_2}+\ldots\text{ },
\label{restrans}
\ee
where the nonlinear terms are admissible only if they satisfy the resonance
conditions
\be
\delta_{mn}=\delta_{m_1n_1}+\ldots+\delta_{m_kn_k}\,.
\label{rescond}
\ee
This is why formula~\eqref{restrans} is sometimes called the resonance
transformation. The idea was to tune the parameters $A_{mn}^{m_1n_1,\dots}$
of this transformation in order to satisfy the basic requirements of the MLG
theory, namely, the conformal selection rules for the correlation numbers
inherited from the minimal model~\cite{Belavin:1984vu} representing its matter sector.\footnote{We
note that because of the structure of the spectrum of the gravitational
dimensions in each particular model, the form of the resonance
transformations is highly constrained and the problem becomes quite
nontrivial (see, e.g.,~\cite{Belavin:2008kv}).}

Nevertheless, it appears that except for the Lee--Yang series
$(2,2s+1)$~\cite{Moore:1991ir, Belavin:2008kv}, this program can be followed
literally only up to two-point correlators. Although there is full agreement
when the fusion rules are satisfied, a discrepancy appears in the
{\it non-physical region}, i.e., when they are not satisfied. For example,
there are three-point correlators, which must be zero according to the
selection rules, but they cannot be made so using the resonance
transformation. It was therefore conjectured that the dual description does
not exactly correspond to the MLG models but describes somewhat modified
theories.

Further progress in understanding the dual models was achieved after the
relation of the Douglas approach to the $(q,p)$ MLG with the $A_{q-1}$
Frobenius manifold structure was revealed (see~\cite{Belavin:2013,VBelavin:2014fs,VBelavin:2014mlg,VBelavin:2014dec}
and the references therein). In particular, it became clear that the
generating function of the correlation numbers is just the tau function of
the Gelfand--Dikij integrable hierarchy connected with the $A_{q-1}$
Frobenius manifold. From this relation, we can derive the nice
representation
\be
Z=\frac{1}{2}\int\limits_0^{\upsilon_*}C_{\alpha\beta\gamma}(v)
\frac{\partial S}{\partial\upsilon_{\beta}}\frac{\partial S}
{\partial\upsilon_{\gamma}}d\upsilon^{\alpha}\,,
\label{Z}
\ee
which will be important for our purposes. Here, $v_\alpha$ ($\alpha=1,\dots,q-1$)
are the flat coordinates on the FM, and $C_{\alpha\beta\gamma}$ are the 
structure constants of the $A_{q-1}$ Frobenius algebra, the algebra of
polynomials modulo the ideal generated by the polynomial $\frac{dQ}{dy}$.
We discuss the properties of the action $S$ in the flat coordinates later.
Perhaps the most important ingredient in~\eqref{Z} is the upper limit $v_*$,
which is a special solution of the string equation
\be
\frac{\partial S}{\partial v_\alpha}\big(v_*\big)= 0\,.
\ee
It was argued in~\cite{VBelavin:2014fs} that only one solution of the
string equation with the special property ${v_{*}}_\alpha(\lambda_{mn})=0$
for $\alpha>1$ and $\lambda_{mn}=0$ (except $\lambda_{11}=\mu$) allows
satisfying the conformal selection rules. After the transition to the flat
coordinates, the necessary expressions for the structure constants were
obtained in~\cite{VBelavin:2014mlg}. All these results made calculations in
the discrete approach very clear both technically and conceptually.

As already mentioned, such formulated theories cannot be regarded as exactly
the same MLG theories, because it is impossible to satisfy the selection
rules for all correlation numbers. The natural question is what can be
modified in order to obtain an exact discrete analogue of the MLG theory.
The first thing that comes to mind is to analyze possible modifications of
the relevant FM structures. Without going into detail, we note that a FM is
a quite rigid construction that, in particular, intrinsically contains a
special Milnor ring (see, e.g.,~\cite{Lee2015}). It is quite natural to try
a possible modification in the simplest case of the Lee--Yang series
$(2,p)$. But in this case, we have $Q'=2y$, and the corresponding algebra is
trivial, $A_1=1$. According to the previous considerations, all physics in
this case is concentrated in the form of resonance
transformation~\eqref{restrans}. In fact, this seems rather strange because
the relation between Milnor rings and Verlinde algebras (fusion rings)
appearing in the conformal field theories makes us think that the
information about selection rules should be encoded in the structure of the
Frobenius algebra itself. For the $(2,p)$ series, for example, the
dimensionality of the possible candidate should correlate with the
dimensionality of the Kac table (or simply with $p$). A question arises
here: Can we use the $A_{p-1}$ algebra to describe the $(2,p)$ series of MLG
models? The first answer is no, because we must construct at least the same
spectrum of gravitational dimensions as we had using $A_1$. But we note that
the spectrum depends not only on the quasihomogeneity property of the
polynomial $Q(y)$ but also on the structure of the action $S$.
In this paper we show that the modification can be made properly such that the
spectrum of the scaling dimensions reproduces the spectrum that appears in
the continuous formulation. Further, we calculate the correlation numbers up
to four-point correlators and show that the results agree perfectly with the
results of the continuous approach without any need for the resonance
transformations.

\section{Calculation of one- and two-point correlation numbers}
\label{sec:FrobMan}

We consider the series of $(2,p)$, ($p\geq5$ and $p$ is odd) minimal models
coupled to Liouville gravity in the spherical topology. The primary fields
are enumerated as $O_{n}=O_{p-n}$, $n\in[1,\frac{p-1}{2}]$. As discussed, we
work with the polynomial $Q(y)=y^p+u_1y^{p-2}+\ldots+u_{p-1}$ instead of
$Q(y)=y^2+u_1$. We first ensure that we obtain spectrum (1.3), i.e.,
$O_k\sim\mu^{-\frac{k+1}{2}}$. The action $S$ with the appropriate scaling
properties is
\be
S=\underset{y=\infty}{\text{res}}
\left(Q^{1+\frac{2}{p}}+\sum_{n=1}^{2n<p}\lambda_{n}Q^{\frac{p-2n}{p}}\right).
\ee
Because $Q^{1+\frac{2}{p}}\sim\mu Q^{\frac{p-2}{p}}$ and
$Q^{1+\frac{2}{p}}\sim\lambda_k Q^{\frac{p-2k}{p}}$, we have
$\lambda_k\sim O_k^{-1}\sim Q^{\frac{2+2k}{p}}\sim\mu^{\frac{k+1}{2}}$.
Using the definition
\be
\theta_{\alpha,k}=\underset{y=\infty}{\text{res}}Q^{k+\frac{\alpha}{p}}(y),
\ee
we can rewrite our action in terms of $\theta_{\alpha,k}$:
\be
S=\theta_{2,1}+\mu\theta_{p-2,0}+
\sum_{n=2}^{{\frac{p-1}{2}}}\lambda_{n}\theta_{p-2n,0}\text{ },
\ee
where $\mu=\lambda_{1}$. In what follows, we use the
proposition~\cite{VBelavin:2014fs}
\be
\left\{\begin{array}{lr}
k \text{ }\text{ even: }\quad \frac{\partial\theta_{\lambda,k}}
{\partial\upsilon_{\alpha}}=\delta_{\lambda,\alpha}x_{\lambda,k}
\left(-\frac{\upsilon_1}{p}\right)^{\frac{k}{2}p},\\
k \text{ }\text{ odd: }\quad \frac{\partial\theta_{\lambda,k}}
{\partial\upsilon_{\alpha}}=\delta_{\lambda,p-\alpha}y_{\lambda,k}
\left(-\frac{\upsilon_1}{p}\right)^{\frac{k-1}{2}p+\lambda},
\end{array}\right.
\ee
where
\be
x_{\lambda,k}=\frac{\Gamma\left( \frac{\lambda}{p}\right)}
{\Gamma\left( \frac{\lambda}{p}+\frac{k}{2} \right)
\left( \frac{k}{2} \right)!}\quad\text{and}\quad y_{\lambda,k}=
-\frac{\Gamma\left( \frac{\lambda}{p}\right)}
{\Gamma\left( \frac{\lambda}{p}+\frac{k+1}{2}\right)
\left( \frac{k-1}{2} \right)!}.
\ee
In the same way as in~\cite{VBelavin:2014dec}, we can obtain the formulas
for the structure constant
\be
C_{\alpha\beta\gamma}=
\left(-\frac{\upsilon_1}{p}\right)^{\frac{\alpha+\beta+\gamma-p-1}{2}}
\theta(\alpha,\beta,\gamma),
\ee
where $\theta(\alpha,\beta,\gamma)=1\Leftrightarrow
\alpha\in[|\beta+\gamma-p|+1:2:p-1-|\beta-\gamma|]$, and its derivative in
the flat coordinates on the line $\upsilon_{\alpha>0}=0$,
\be
\partial_{\delta}C_{\alpha\beta\gamma}=\theta(\alpha,\beta,\gamma,\delta,p)
\frac{2p-\alpha-\beta-\gamma-\rho}{2p}
\left(-\frac{\upsilon_1}{p}\right)^{\frac{\alpha+\beta+\gamma+\rho-2p-2}{2}},
\ee
if $(\alpha+\beta+\gamma+\rho-2p-2)/2\in\mathbb{N}_0$, where
\begin{align}
\theta(\alpha,\beta,\gamma,\delta,p)=&[(p-m_1)
\chi_{1,m_1}(m_2+m_3-m_4)+\frac{2p+m_4-m_1-m_2-m_3}{2}\times
\nonumber\\
&\times\chi_{m_1+2,2p-m_1-2}(m_2+m_3-m_4)]\,,
\end{align}
and
$m_i=\mathrm{RankedMax}[\{\alpha,\beta,\gamma,\delta\},i]$.

We now calculate one-point numbers. As we know from the conformal selection
rules, the one-point correlation numbers of all operators except the unity
operator must be zero. Indeed, we find
\be
Z_1=\langle O_{n}\rangle=\int\limits_0^{\upsilon_{*1}}C_{p-1,\alpha,\beta}
\frac{\partial S^{(0)}}{\partial\upsilon_{\alpha}}
\frac{\partial S^{(n)}}{\partial\upsilon_{\beta}}d\upsilon_1.
\ee
Taking into account that
\be
C_{p-1,\alpha,\beta}\sim\delta_{\alpha,\beta},\quad
\frac{\partial S^{(0)}}{\partial\upsilon_{\alpha}}\sim
\delta_{\alpha,p-2},\quad
\frac{\partial S^{(n)}}{\partial\upsilon_{\alpha}}\sim\delta_{\alpha,p-2n},
\ee
it follows that the one-point function $Z_1=0$ for $n\neq1$. In particular,
in the case $n=1$, we find
\be
Z_{0}=\langle O_{1}\rangle=\frac{1}{2}\int\limits_0^{\upsilon_{*1}}
C_{p-1,\alpha,\beta}\frac{\partial S^{(0)}}{\partial\upsilon_{\alpha}}
\frac{\partial S^{(0)}}{\partial\upsilon_{\beta}}d\upsilon_1=
\frac{1}{2}\int\limits_0^{\upsilon_{*1}}C_{p-1,p-2,p-2}
\left(y_{2,1}\left(\frac{\upsilon_1}{p}\right)^2+\mu\right)^2d\upsilon_1.
\ee
Substituting $\mu=\frac{\upsilon_{*1}^2}{2p}$, we can obtain
\be
Z_0=\frac{1}{(p-2)p(p+2)}\frac{{\upsilon_{*1}}^{p+2}}{p^{p+3}}.
\ee
And the two-point correlator is
\be
Z_{12}=\langle O_{n_1}O_{n_2}\rangle=\int\limits_0^{\upsilon_{*1}}
C_{p-1,\alpha,\beta}\frac{\partial S^{(n_1)}}{\partial\upsilon_{\alpha}}
\frac{\partial S^{(n_2)}}{\partial\upsilon_{\beta}}d\upsilon_1=
%\delta_{n_1,n_2}\int\limits_0^{\upsilon_{*1}}\left(-\frac{\upsilon_1}{p}\right)^{p-2n_1-1}d\upsilon_1%
\delta_{n_1,n_2}\left(-\frac{1}{p}\right)^{p-2n_1-1}
\frac{{\upsilon_{*1}}^{p-2n_1}}{p-2n_1}.
\ee
Hence, the two-point correlators have a proper diagonal form. On the other
hand, a simple analysis based on the new action $S$ shows that there is no
way to use the freedom of the resonance transformations in this case.

\section{Calculation of three-point correlators}
\label{sec:SolutionPlan}

In this section, we calculate three-point numbers and compare the resulting
normalized expression with the expression from the continuous approach,
\be
Z_{123}=\langle O_{n_1}O_{n_2}O_{n_3}\rangle=C^{\alpha\beta}_{\gamma}
\frac{\partial\upsilon_*^{\gamma}}{\partial \lambda_{n_3}}
\frac{\partial S^{(n_1)}}{\partial\upsilon^{\alpha}}
\frac{\partial S^{(n_2)}}{\partial\upsilon^{\beta}}=C_{\gamma,p-2n_1,p-2n_2}
\frac{\partial\upsilon_*^{\gamma}}{\partial \lambda_{n_3}}.
\ee
To obtain an expression for $\frac{\partial\upsilon_*^{\gamma}}
{\partial \lambda_{n_3}}$, we proceed as follows. We use the Douglas string
equation as
\be
\left.\frac{\partial}{\partial \lambda_k}
\frac{\partial S}{\partial\upsilon^{\alpha}}\right|_{\upsilon_*}=0,\quad
\alpha=1,\ldots,p-1.
\ee
Substituting (2.3), we have
\be
\frac{\partial}{\partial\lambda_k}\left[\frac{\partial\theta_{2,1}}
{\partial\upsilon^{\alpha}}+\mu \frac{\partial\theta_{p-2,0}}
{\partial\upsilon^{\alpha}} +\sum_{n=2}^{\frac{p-1}{2}}\lambda_n
\frac{\partial\theta_{p-2n,0}}{\partial\upsilon^{\alpha}} \right]=0.
\ee
Hence, we obtain the following useful equation
\be
C_{2\alpha\beta}\frac{\partial\upsilon_*^{\beta}}
{\partial \lambda_k}+\delta_{\alpha,2k}=0.
\ee
Taking the special nonzero condition for $C_{2\alpha\beta}$ into account,
namely, $C_{2\alpha\beta}\neq0$ iff $\alpha+\beta=p\pm1$, we can recursively
obtain an expression for $\frac{\partial\upsilon^{\beta}_*}
{\partial \lambda_k}$. Indeed, considering (3.4) for $\alpha=1$ and for
$\alpha=p-1$, we obtain
\be
\frac{\partial\upsilon^{p-2}_*}{\partial \lambda_k}=0,\quad
\frac{\partial\upsilon^{2}_*}{\partial \lambda_k}=
\frac{p}{\upsilon_{*1}}\delta_{k,\frac{p-1}{2}}.
\ee
If $\alpha\neq p-1$ and $\alpha\neq1$, then $\beta=p+1-\alpha$ or
$\beta=p-1-\alpha$, and we obtain
\be
\frac{\partial\upsilon_*^{p+1-\alpha}}{\partial\lambda_k}=
\frac{p}{\upsilon_{*1}}\left( \frac{\partial\upsilon_*^{p-1-\alpha}}
{\partial \lambda_k} +\delta_{\alpha,2k}\right)
\ee
from (3.4). By recursive computation, we can get zero for odd $\beta$ and
for even $\beta$:
\be
\frac{\partial\upsilon^{\beta}_*}{\partial \lambda_k}=
\left(\frac{p}{\upsilon_{*1}}\right)^{\frac{\beta+2k+1-p}{2}}\quad\text{if}
\quad\frac{\beta}{2}\geq \frac{p+1}{2}-k.
\ee
Combining (2.6) and (3.7), we obtain
\be
Z_{123}=\left( \frac{\upsilon_{*1}}{p} \right)^{p-1-\sum_{i=1}^3n_i}
\sum_{n=\max\{1,\frac{p+1}{2}-n_3\}}^{\frac{p-1}{2}}
(-1)^{\frac{p-1}{2}-n_1-n_2+n}\theta(2n,p-2n_1,p-2n_2).
\ee
After some operations with this expression, we finally find
\be
Z_{123}=\left(\frac{\upsilon_{*1}}{p}\right)^{p-1-\sum_{i=1}^3n_i}\theta_{123},
\ee
where $\theta_{123}$ denotes the nonzero condition of this expression. It
turns out that it coincides with the selection rules for three-point
correlators,\footnote{In the next condition, there is no minimum in the
upper limits as soon as we take $n_i\in[1,\frac{p-1}{2}]$.} which come from
the CFT fusion rules, i.e., $\theta_{123}=1$ if
$n_3\in[|n_1-n_2|+1:2:n_1+n_2-1]$ or $p-n_3\in[|n_1-n_2|+1:2:n_1+n_2-1]$.

We now have all necessary ingredients for comparing with the continuous
approach. Using (2.12), (2.13), and (3.9), we find the normalized expression
for three-point numbers
\be
\frac{Z_{123}^2Z_0}{Z_{11}Z_{22}Z_{33}}=
\frac{\prod_{i=1}^3(p-2n_i)}{(p-2)p(p+2)}\theta_{123}.
\ee
This is exactly the same as the continuous expression calculated
in~\cite{Zam:2005} for general $(q,p)$ models
\be
\frac{\langle\langle O_{m_1n_2}O_{m_2n_2}O_{m_3n_3}\rangle\rangle^2}
{\prod_{i=1}^3\langle\langle O^2_{m_in_i}\rangle\rangle}=
\frac{\prod_{i=1}^3|m_ip-n_iq|}{p(p+q)(p-q)}\theta^{pq}_{123},
\ee
where $\theta^{pq}_{123}$ denotes the selection rules of general $(q,p)$
models.

\section{Calculation of four-point correlators}
\label{sec:AppSol}

In this section, we calculate four-point numbers and compare their
normalized expression with the expressions found in the continuous approach:
$$
Z^{disc}_{1234}=\frac{\partial^2\upsilon_{*}^{\gamma}}
{\partial \lambda_3\partial \lambda_4}C_{\gamma}^{\alpha\beta}
\frac{\partial S^{(n_1)}}{\partial\upsilon^{\alpha}}
\frac{\partial S^{(n_2)}}{\partial\upsilon^{\beta}}+
\frac{\partial\upsilon_{*}^{\gamma}}{\partial \lambda_3}
\frac{\partial C_{\gamma}^{\alpha\beta}}{\partial \lambda_4}
\frac{\partial S^{(n_1)}}{\partial\upsilon^{\alpha}}
\frac{\partial S^{(n_2)}}{\partial\upsilon^{\beta}}=
$$
\be
=\frac{\partial^2\upsilon_{*}^{\gamma}}{\partial\lambda_3\partial\lambda_4}
C_{\gamma,p-2n_1,p-2n_2}+\partial _{\delta}C_{\gamma,p-2n_1,p-2n_2}
\frac{\partial\upsilon_{*}^{\gamma}}{\partial \lambda_3}
\frac{\partial\upsilon_*^{\delta}}{\partial \lambda_4}.
\ee
To obtain an expression for $\frac{\partial^2\upsilon_{*}^{\gamma}}
{\partial\lambda_3\partial\lambda_4}$, we proceed the same way as for
calculating $\frac{\partial\upsilon^{\beta}_*}{\partial \lambda_k}$.
Differentiating (3.4) with respect to $\lambda_j$, we obtain
\be
C_{2\alpha\beta}\frac{\partial^2\upsilon_{*}^{\beta}}
{\partial \lambda_j\partial \lambda_k}+\partial_{\gamma}C_{2\alpha\beta}
\frac{\partial\upsilon_{*}^{\gamma}}{\partial \lambda_j}
\frac{\partial\upsilon_{*}^{\beta}}{\partial \lambda_k}=0.
\ee
Noting that $\partial_{\gamma}C_{2\alpha\beta}=-\frac{1}{p}
\delta_{\alpha+\beta+\gamma,2p}$ , we rewrite (4.2) as
\be
C_{2\alpha\beta}\frac{\partial^2\upsilon_{*}^{\beta}}
{\partial \lambda_j\partial \lambda_k}-\frac{1}{p}
\left( \frac{p}{\upsilon_{*1}} \right)^{n_j+n_k-\frac{p-1}{2}}
\sum_{n=\frac{p+1}{2}-n_j}^{\frac{p-1}{2}}
\sum_{m=\frac{p+1}{2}-n_k}^{\frac{p-1}{2}}
\left( \frac{p}{\upsilon_{*1}} \right)^{n+m}\delta_{\alpha,2p-2n-2m}.
\ee
Using the ansatz
\be
\frac{\partial^2\upsilon_{*}^{\beta}}{\partial \lambda_j\partial \lambda_k}=
-\frac{1}{p}\left( \frac{p}{\upsilon_{*1}} \right)^{n_j+n_k-\frac{p-1}{2}}
\sum_{n=\frac{p+1}{2}-n_j}^{\frac{p-1}{2}}
\sum_{m=\frac{p+1}{2}-n_k}^{\frac{p-1}{2}}f(\beta,p-n-m),
\ee
in the same way as we obtained result (3.7), we obtain
\be
\frac{\partial^2\upsilon_{*}^{\beta}}{\partial \lambda_j\partial \lambda_k}=
-\frac{1}{p}\left( \frac{p}{\upsilon_{*1}} \right)^{n_j+n_k-\frac{p-3}{2}+
\frac{\beta}{2}}\sum_{n=\frac{p+1}{2}-n_j}^{\frac{p-1}{2}}
\sum_{m=\frac{p+1}{2}-n_k}^{\frac{p-1}{2}}
\sum_{i=1}^{\beta/2}\delta_{n+m,\frac{p-1}{2}+i}
\ee
for even $\beta$. Combining (2.6) and (4.5), we can see that the first term
in (4.1) is
\be
\frac{\partial^2\upsilon_{*}^{\gamma}}{\partial\lambda_3\partial\lambda_4}
C_{\gamma,p-2n_1,p-2n_2}=\frac{1}{p}
\left( \frac{p}{\upsilon_{*1} }\right)^{\sum_{i=1}^4n_i+2-p}
\sum_{t=|\frac{p}{2}-n_1-n_2|+\frac{1}{2}}^{\frac{p-1}{2}-|n_1-n_2|}
(-1)^{\frac{p+1}{2}-n_1-n_2+t}\sum_{i=1}^t\varphi(i),
\ee
where
\be
\varphi(i)=\sum_{n=\frac{p+1}{2}-n_3}^{\frac{p-1}{2}}
\sum_{m=\frac{p+1}{2}-n_4}^{\frac{p-1}{2}}\delta_{m+n,\frac{p-1}{2}+i}
\ee
or, explicitly,
\be
\varphi(i)=\frac{p+1-2i}{4}+\frac{|n_3+n_4-\frac{p+1}{2}+i|}{2}-
\frac{|n_3-\frac{p+1}{2}+i|}{2}-\frac{|n_4-\frac{p+1}{2}+i|}{2}.
\ee
Using (2.7), (3.7), and (4.6), we obtain the normalized expression
$$
\frac{Z^{disc}_{1234}Z_0}{\sqrt{Z_{11}Z_{22}Z_{33}Z_{44}}}=
\frac{\sqrt{\prod_i(p-2n_i)}}{(p-2)p(p+2)}\times
$$
\be
\times\left[\sum_{t=|\frac{p}{2}-n_1-n_2|+\frac{1}{2}}^{\frac{p-1}{2}-|n_1-n_2|}
(-1)^{\frac{p+1}{2}-n_1-n_2+t}\sum_{i=1}^t\varphi(i)+
\sum_{n=\frac{p+1}{2}-n_3}^{\frac{p-1}{2}}
\sum_{m=\frac{p+1}{2}-n_4}^{\frac{p-1}{2}}\mathrm{F}(n_1,n_2,n,m,p)\right],
\label{Z1234disc}
\ee
where
\be
\mathrm{F}(n_1,n_2,n,m,p)=(m+n-n_1-n_2)(-1)^{n+m-n_1-n_2}\theta(2n,2m,p-2n_1,p-2n_2).
\ee
It is easy to verify that \eqref{Z1234disc} is symmetric. In the continuous approach,\\
 for $n_1\leq\ldots\leq n_4$, this quantity is
\be
\frac{Z^{cont}_{1234}Z_0}{\sqrt{Z_{11}Z_{22}Z_{33}Z_{44}}}=
\frac{\sqrt{\prod_i(p-2n_i)}}{2(p-2)p(p+2)}
\left(\sum_{i=2}^4\sum_{t=-(n_1-1)}^{n_1-1}|p-2(n_i-t)|-n_1(p+2n_1)\right).
\label{Z1234cont}
\ee
We recall that \eqref{Z1234cont} is obtained~\cite{Belavin:2006kv} under the
assumption that the number of conformal blocks of the four-point correlator
is exactly $n_1$. This can be expressed through the following condition
\be
\left\{
\begin{array}{lr}
n_1+n_2+n_3+n_4\text{ }\text{ even: } \quad n_1+n_4\leq n_2+n_3,\\
n_1+n_2+n_3+n_4\text{ }\text{ odd: }\quad -n_1+n_2+n_3+n_4\geq p-2
\end{array}
\right.
\label{region}
\ee
(which in turn ensures that the selection rules are satisfied). We find that
in this region, \eqref{Z1234disc} coincides with \eqref{Z1234cont}, while
outside of this region our expression (4.9) gives zero values. It is
interesting that the previous consideration based on the polynomial $Q$ of
the second degree in $y$ gives sometimes nonzero values outside~\eqref{region}
\cite{Belavin:2013}, which makes these two results significantly different.
Unfortunately, corresponding results derived in the continuous approach are
currently unknown, and we cannot conclude which of these two discrete
versions is appropriate.

\section{Conclusion}
We have considered another description of the Lee--Yang series of MLG
models. At the level of the correlation functions up to four points, we
verified that the results obtained using this description agree with the
results obtained using the initial continuous definition of MLG.
The computation differs significantly from the
previous one~\cite{Belavin:2008kv,{Moore:1991ir},Belavin:2013}. In
particular, it does not require any resonance transformation. The essential
modification is related to another choice of the FM relevant for the $(q,p)$
MLG model. The construction for the $(2,p)$ model is now based on the
$A_{p-1}$ Frobenius algebra. We find this description more natural because
the dimension of the $A_{p-1}$ algebra exactly equals the number of basic
physical BRST cohomologies constructed from the primary fields in the
minimal model $(2,p)$ (modulo the standard symmetry factor 2 of the Kac
table). It can therefore play the role of the regulator ensuring the
necessary satisfaction of the selection rules, while in the previous 
scheme this role had to be solved by supplementary resonance transformations.

It would be interesting to seek a possible alternative description without
resonances in the general $(q,p)$ case. Certainly, this question is related
to the longstanding problem of $p{-}q$ duality in the discrete approach to
MLG~\cite{Ginsparg:1990zc,Fukuma1992}. Apparently, an appropriate description should be
symmetric under the interchange of $q$ and $p$.

Another interesting issue is related to the problem described at the end of
the preceding section: the origin of the discrepancy between the two
versions in the region outside~\eqref{region}. The question of computing
(spherical and higher-genera) multipoint correlators (and also the
correlators of the gravitational descendants) deserves to be studied in this
perspective. This problem, in particular, requires more detailed analysis of the
structure constants of the $A_{p-1}$ Frobenius algebra in the flat
coordinates. We plan to investigate this question in the near future.

\vspace{5mm}

\noindent \textbf{Acknowledgments.}
We are grateful to A.~Belavin for reading the manuscript and making useful
comments. The work was performed at the Institute for Information
Transmission Problems with the financial support of the Russian Science
Foundation (Grant No.14-50-00150).

\vspace{5mm}

%%%%%%%%%%%%%%%%%%%%%%%%%%%%%%%%%%%%%%%%%%%%%%%%%%%%%%%%%%%%%%%%%%%%%%%%
%%%%%%%%%%%%%%%%%%%%%%%%%%%%%%%%%%%%%%%%%%%%%%%%%%%%%%%%%%%%%%%%%%%%%%%%
%%%%%%%%%%%%%%%%%%%%%%%%%%%%%%%%%%%%%%%%%%%%%%%%%%%%%%%%%%%%%%%%%%%%%%%%

\providecommand{\href}[2]{#2}\begingroup\raggedright
\addtolength{\baselineskip}{-3pt} \addtolength{\parskip}{-1pt}

\end{document}